\documentclass[journal]{IEEEtran}
\usepackage{graphicx}
\usepackage{amsmath}
\usepackage{url}
\usepackage{subcaption}

\ifCLASSINFOpdf
\fi
\begin{document}
\title{Nonvolatile Memory Cell Based on Memristor Emulator}
\author{Santosh Parajuli, Ram Kaji Budhathoki, Hyongsuk Kim ~\IEEEmembership{}
\thanks{Santosh Parajuli and Ram Kaji Budhathoki are with the Department
of Electrical and Electronics Engineering, Kathmandu University, Nepal, e-mail: (eesantosh@ku.edu.np, ram.budhathoki@ku.edu.np). Hyongsuk Kim is with the  Division of Electronics and Information Engineering, Chonbuk National University, Jeonju 567-54896, South Korea, e-mail:(hskim@jbnu.ac.kr).}}

\markboth{}%
{Shell \MakeLowercase{\textit{et al.}}: Bare Demo of IEEEtran.cls for Journals}
\maketitle
\begin{abstract}
Memristor, one of the fundamental circuit elements, has promising applications in non-volatile memory and storage technology as it can theoretically achieve infinite states. Information can be stored independently in these states and retrieved whenever required. In this paper, we have proposed a non volatile memory cell based on memristor emulator. The circuit is able to perform read and write operations. In this memristor based memroy cell, unipolar pulse is used for writing and bipolar pulse is used for reading. Unlike other earlier designs, the circuit does not need external read/write enable switches to switch between read and write operations; the switching is achieved by the zero average bipolar read pulse given after the completion of write cycle. In our proposed memristor based memory cell, single bit can be read and any voltages from 0 to 5 volts can be written. Mathematical analysis and the simulation results of memristor emulator based read write circuit have been presented to confirm its operation.            
\end{abstract}

\IEEEpeerreviewmaketitle

\section{Introduction}




\IEEEPARstart{M}{emristor} was first postulated by Leon Chua in 1971 as a fundamental circuit element \cite{chuamem}. In a memristor when the current flows in  one direction, it's resistance decreases and vice versa. When the current flow is stopped, memristor retains its final state. This state retention property of memristor makes it useful as a non-volatile memory element \cite{helen}, \cite{stanley}.

Memristor is a natural application for resistive random access memory (ReRAM) technology \cite{jjyangnature}; moreover, neuromorphic computation using ReRAM technology is also becoming popular \cite{helen}, \cite{yeonjoo}, \cite{miaohu}, \cite{xiao}.
By controlling programming signal width and/or amplitude, memristor can be taken to any state; however, two  states: high resistance state (HRS) and low resistance state (LRS) are commonly used and they represent bit 1 and bit 0 respectively \cite{helen},
\cite{yanglu}, \cite{yenpo}, \cite{elshamy}. 

Several research groups have proposed memristor emulators \cite{hkim}, \cite{lopez}, \cite{sozen},  \cite{yesil}, \cite{yud}. Lopez \textit{et al.}  \cite{lopez} and Yu \textit{et al.} \cite{yud} have proposed a floating memristor emulator circuit based on current conveyors. The emulator proposed by Kim \textit{et al.} \cite{hkim} is built from off-the-shelf solid state components. In this paper, \cite{hkim} has been chosen for emulating a memristor because it
has shown promising results that provides an alternative solution of \textit{hp} TiO$_2$ memristor model in real circuit. 

Memristor memory cell architecture based on the memristor model has been proposed by \cite{yenpo}, \cite{elshamy}, \cite{bmohammad}, \cite{sarwar}. 
Mohammad \textit{et al.} \cite{bmohammad} have explored
various aspects of memristor modeling and designed a memristor based memory. Likewise, Sarwar \textit{et al.} \cite{sarwar} has designed memristor based nonvolatile random access memory, but the memristor in their work uses Spice model. The design analysis of nonvolatile memristor memories, \cite{yenpo}, has been used for designing a read write circuit in this paper because their analysis have specifically targeted key electrical memristor characteristics relevant to memory operations. 
 
Even though many researchers have proposed memristor emulator circuit and memory cell based on memristor model, \emph{no memristor emulator based read write circuit is available}, which can correctly estimate the behavior of a real physical memristor memory cell. In \cite{yenpo}, memristor linear model is used for realizing a memory element. In this paper memristor emulator is used for realizing a memory cell. Furthermore, the need for external read write enable switch in \cite{yenpo} and \cite{elshamy} is eliminated. The unipolar write pulse and bipolar read pulse with redefined read circuit helps to eliminate the read write enable switch. This not only completely removes switching loss in read write enable switch but also reduces the access time. Due to read cycle distortion, the memristor internal state may change and enter into the invalid state. This read distortion issue of the memristor memory is also investigated. Read access time duration in read pattern is proposed to preserve data integrity during the read operation.

The memristor state in this work is assumed to be the voltage across capacitor of the memristor emulator circuit; in fact, actual memristor state is obtained by  multiplying instantaneous capacitor voltage by a fixed resistor and adding another fixed resistor as shown in \cite{hkim}.
Memristor state, memristor resistance, memristance are used interchangeably throughout the paper. Rest of the paper is organized as follows: section II gives brief introduction of memristor linear model followed by the discussion of Kim's emulator. Section III discusses the proposed read write circuit and its operation. Simulation results of our memristor emulator based read write circuit is presented in section IV. Finally, a short conclusion is given in section V.

\section{Principle of Memristor and Memristor Emulator Circuit}
From a circuit perspective, memristor is often modeled as a two terminal device, with two electrodes and a sandwiched conductive channel which acts as a switching layer \cite{stanley}, \cite{yang}. 
In general the memristive systems are defined as \cite{kang}  
\begin{equation}
\label{Eq: 1}
\frac{dx}{dt}= f(x,u,t)
\end{equation}
\begin{equation}
\label{Eq: 2}
y=g(x,u,t)u
\end{equation}
where $u$ and $y$ denote the input and output of the system and $x$ denotes the state of the system. From \eqref{Eq: 1}, it can be seen that, in a memristive system the state evolves with time $t$. The equations can be modified by choosing appropriate variables to define memristor for electronic circuit as
\begin{equation}
\label{Eq: 1modified}
\frac{dw}{dt} = f(w,v_M,t)
\end{equation}
\begin{equation}
\label{Eq: 2modified}
i_M=M(w,v_M,t)v_M
\end{equation}
where $v_M$ and $i_M$ denote the current and voltage of the memristor and $w$ denotes the internal state variable. $M$ denotes the memconductance (memristance). As shown in \eqref{Eq: 1modified} -- the voltage input directly affects the state variable $w$ -- which is typical in first order memristive system \cite{yeonjoo}.

\begin{figure}[t]
\center
\includegraphics[scale=0.5]{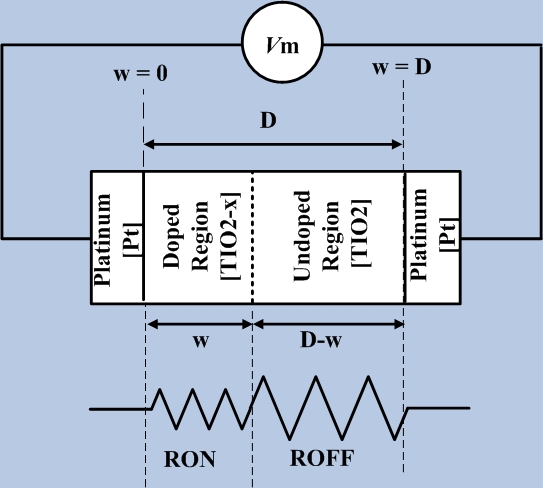}
\caption{\textbf{Linear Memristor Model.}}
\label{Fig: Strukov}
\end{figure}

Memristor linear model shows how memristance arises naturally in nanoscale system due to coupling of electronic and ionic transport under an external bias. Equivalent circuit for this model is shown in Fig. \ref{Fig: Strukov}. Two state dependent resistors are in series, and they represent ON and OFF channel resistance respectively.
The device dynamics and the \textit{i-v} characteristics are described by (\ref{Eq: 4}) and (\ref{Eq: 5}) respectively.
\begin{equation}
\label{Eq: 4}
\frac{dw}{dt} = \mu_v \frac{R_{ON}}{D}i_M
\end{equation}
\begin{equation}
\label{Eq: 5}
v_M = \left(R_{ON}\frac{w}{D}+R_{OFF}\left(1-\frac{w}{D}\right)\right )i_M
\end{equation}
where $\mu_v$ is average ion mobility, $R_{ON}$ is the resistance of doped region, $R_{OFF}$ is the resistance of undoped region,  and $w \in (0,D)$. This model considers ohmic electronic conduction and linear ionic drift, and assumes modulation of $w$ is due to drift of charged dopants \cite{stanley}. 

\begin{figure}[t]
\begin{subfigure}{.5\textwidth}
  \centering
  \includegraphics[width=\linewidth]{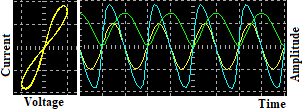}  
  \caption{Pinched hysteresis loop at 100 Hz and corresponding time domain waveform.}
  \label{fig:3a}
\end{subfigure}
\hfill
\begin{subfigure}{.5\textwidth}
  \centering
  \includegraphics[width=\linewidth]{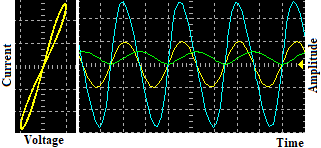}  
  \caption{Pinched hysteresis loop at 400 Hz and corresponding time domain waveform.}
  \label{fig:3b}
\end{subfigure}
\caption{\textbf{Response of Memristor Emulator in XY and YT Mode of an Oscilloscpoe.} In the time domain, yellow trace represents input voltage, green trace represents voltage across capacitor, and blue trace represents waveform generated by current through memristor. Time axis is at 5 mS/div. In YT mode, yellow trace is at 1 V/div, blue trace is at 50 mV/div, and green trace is at 2 V/div.}
\label{Fig: memristorverify}
\end{figure}

The incremental configuration of memristor emulator \cite {hkim} is used in this work. One of the distinguished features of Kim's emulator is that the memristor can be programmed by giving pulse input and the programming information can be stored in the capacitor for longer period of time. This feature of memristor is important to use memristor as a computational memory element \cite{stanford}.  
To confirm the basic memristor properties, 1 volt sinusoidal waveform is given at frequencies of 100 Hz and 400 Hz respectively. Current probe with voltage to current ratio of 1 V/mA
is used for plotting the current trace in oscilloscope. Oscilloscope traces are shown in Fig. \ref{Fig: memristorverify}. Increased height with increment in frequency in the pinched loop confirms incremental configuration of the memristor. In the time domain, green trace is voltage variation across the capacitor, and this voltage is considered as memristor state in our work. 

\begin{figure*}[t]
\begin{subfigure}{.5\textwidth}
  \centering
  \includegraphics[width=\linewidth]{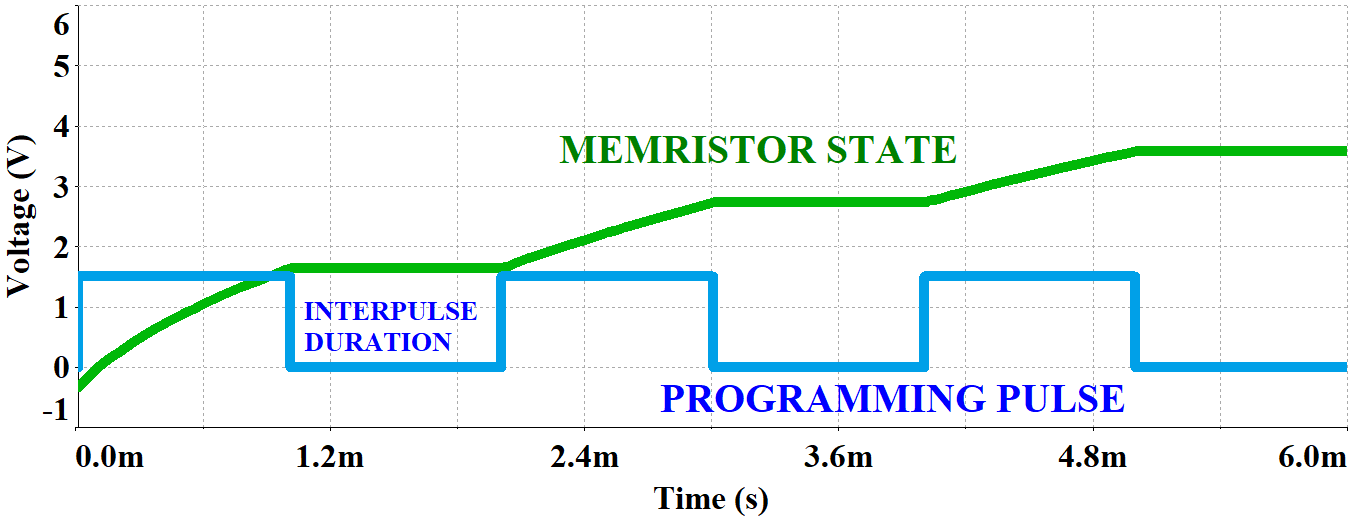}  
  \caption{Change in memristor state with 1.5 volts amplitude programming pulse. The programming pulse has equal ON and OFF time duration.}
  \label{fig:3a}
\end{subfigure}
\hfill
\begin{subfigure}{.5\textwidth}
  \centering
  \includegraphics[width=\linewidth]{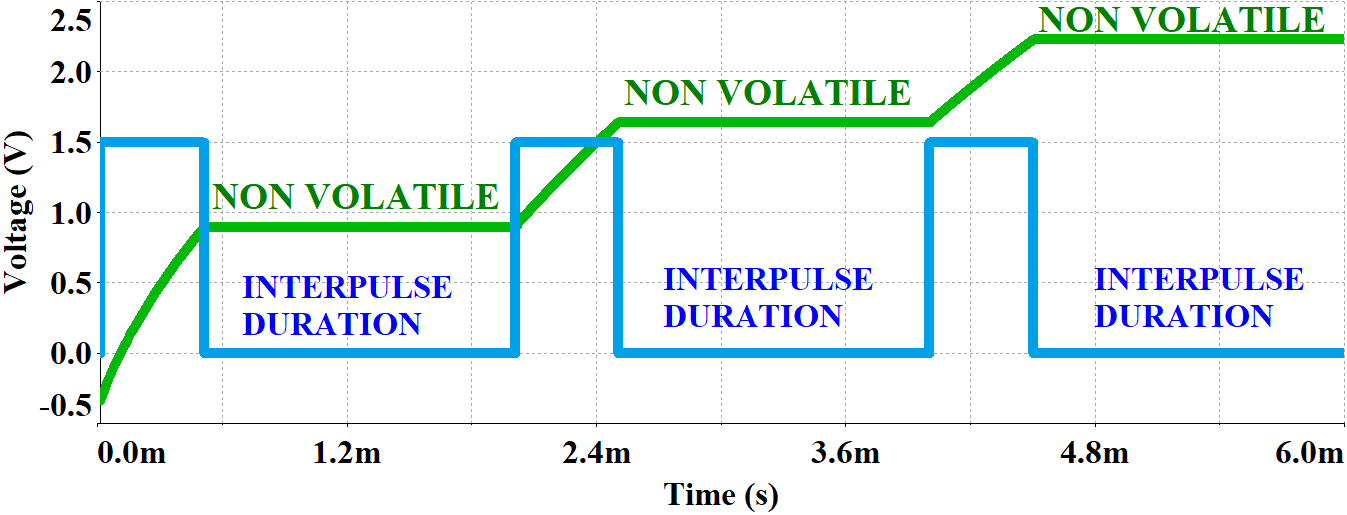}  
  \caption{Change in memristor state with 1.5 volts amplitude programming pulse. The programming pulse has  ON time less than OFF time.}
  \label{fig:3b}
\end{subfigure}

\begin{subfigure}{.5\textwidth}
  \centering
  \includegraphics[width=\linewidth]{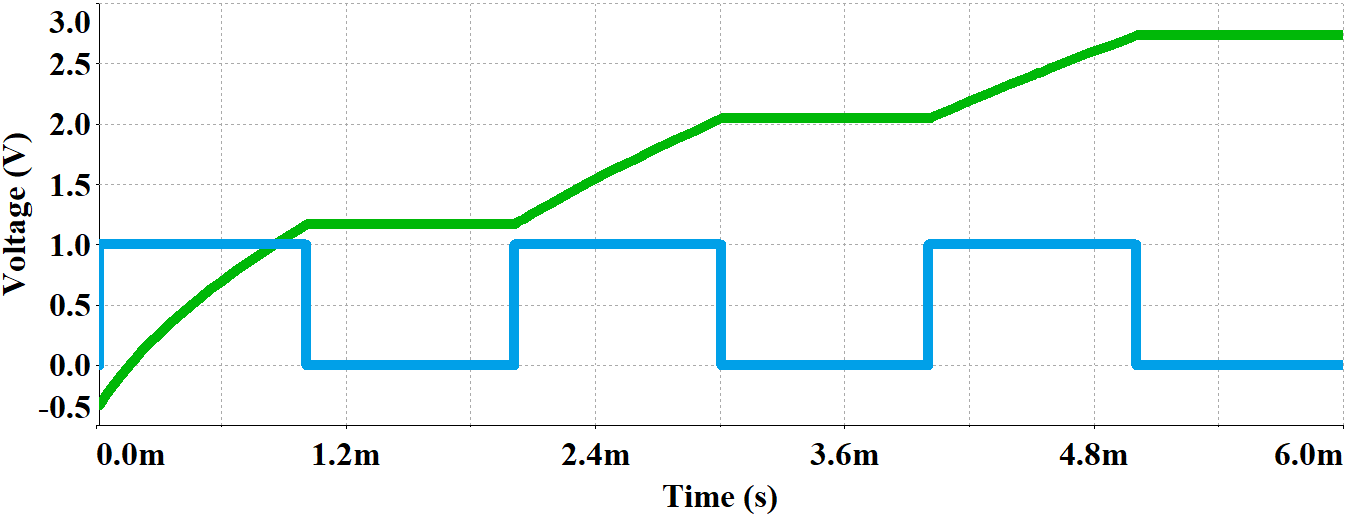}  
  \caption{Change in memristor state with 1 volt amplitude programming pulse. The programming pulse has equal ON and OFF time duration.}
  \label{fig:3c}
\end{subfigure}
\hfill
\begin{subfigure}{.5\textwidth}
  \centering
  \includegraphics[width=\linewidth]{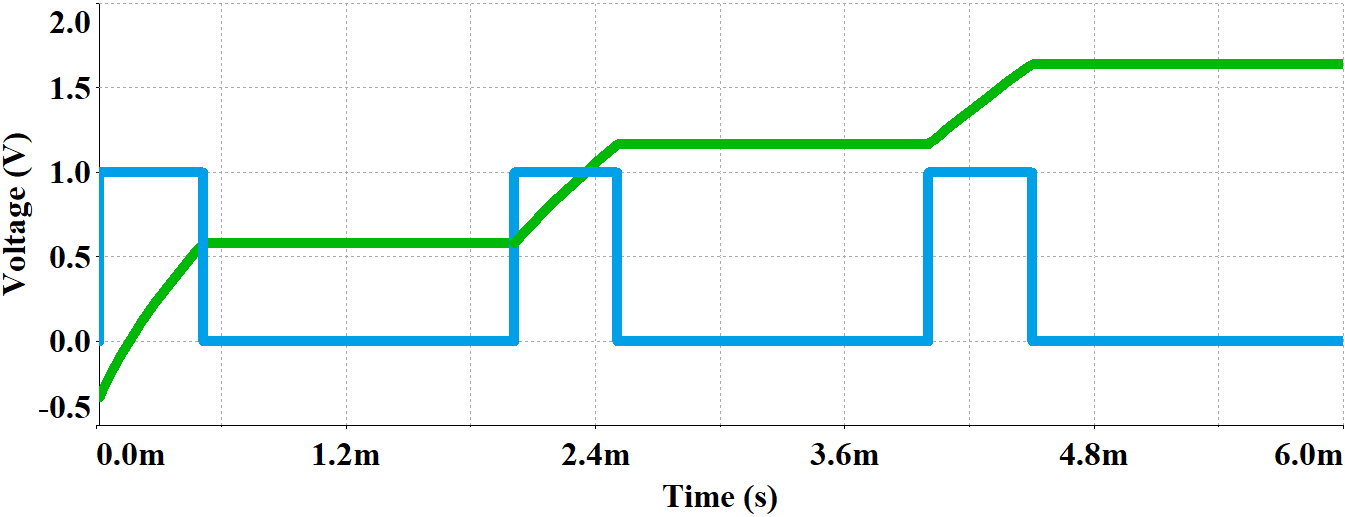}  
  \caption{Change in memristor state with 1 volt amplitude programming pulse. The programming pulse has  ON time less than OFF time.}
  \label{fig:3d}
\end{subfigure}

\caption{\textbf{Effect of Variation of Height and Width of the Programming Pulse on the State of the Memristor.}}
\label{Fig: heightwidth_tile}
\end{figure*}

The response of memristor with programming pulse input is shown in Fig. \ref{Fig: heightwidth_tile}. In all the four cases (Fig. \ref{fig:3a}, Fig. \ref{fig:3b}, Fig. \ref{fig:3c}, Fig. \ref{fig:3d}), the increment step size is smaller as more pulses are applied. This is because, in a memristor present state is the cumulative effect of past applied pulses. This observation leads memristor to be useful as a computational memory element particularly for neuromorphic computing. Non volatility of the memristor state during inter pulse period can also be seen.

The results in Fig. \ref{Fig: memristorverify} and  \ref{Fig: heightwidth_tile} are consistent with the results obtained in \cite{hkim}.



%
%
%
%

\section{Proposed Read Write Circuit}
\begin{figure}[t]
\center
\includegraphics[scale=.26]{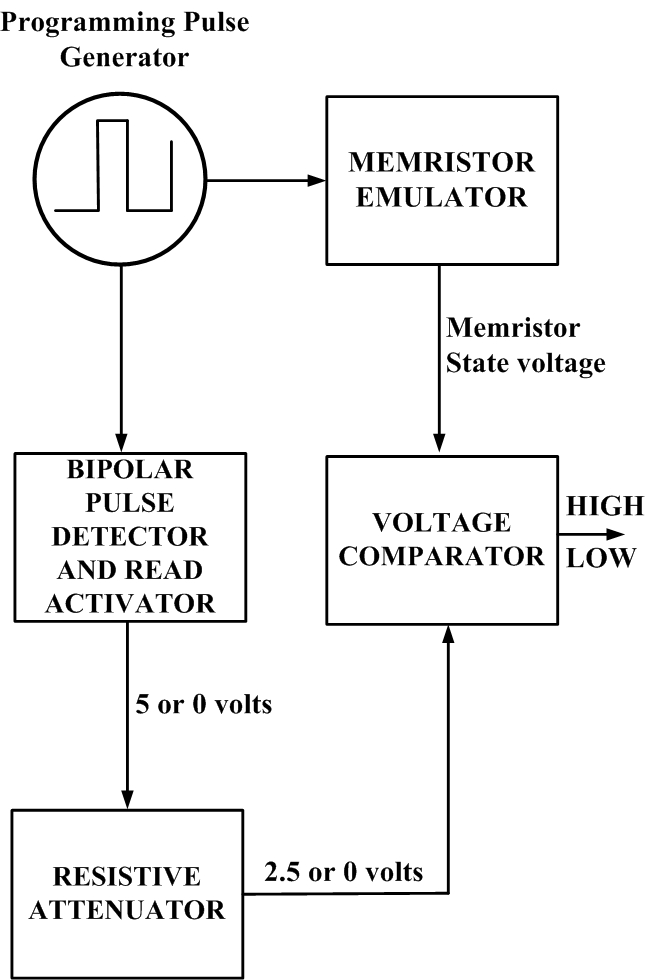}
\caption{\textbf{Block Diagram of the Proposed Read Write Circuit.}}
\label{Fig: signalflow}
\end{figure}

Based on memristor response to pulse input, a read write circuit is proposed in this paper. The block diagram of the read write circuit is  shown in Fig. \ref{Fig: signalflow}. The comparator is used for detecting the state of the memristor (high or low) depending on the nature of programming pulse. The maximum and minimum voltages that can be written in the memristor are 5 volts and 0 volt respectively. 
The read circuit should read high/low, after the read pulse is given. When the memristor state voltage is above 2.5 volts, the read circuit should read high (5 volts). Similarly, when it is below 2.5 volts, the read circuit should read low (0 volt). The read circuit should perform two tasks: first, it should detect read pulse and activate read circuit; second, it should compare the output voltage of memristor with 2.5 volts to detect correct memristor state (high or low). If there is no read pulse, the read circuit should not come into action, and the read circuit should retain it's original state. Any bipolar pulse with zero average value could be chosen to read from memristor without perturbing its internal state.

\begin{figure}[t]
\center
\includegraphics[width=\linewidth]{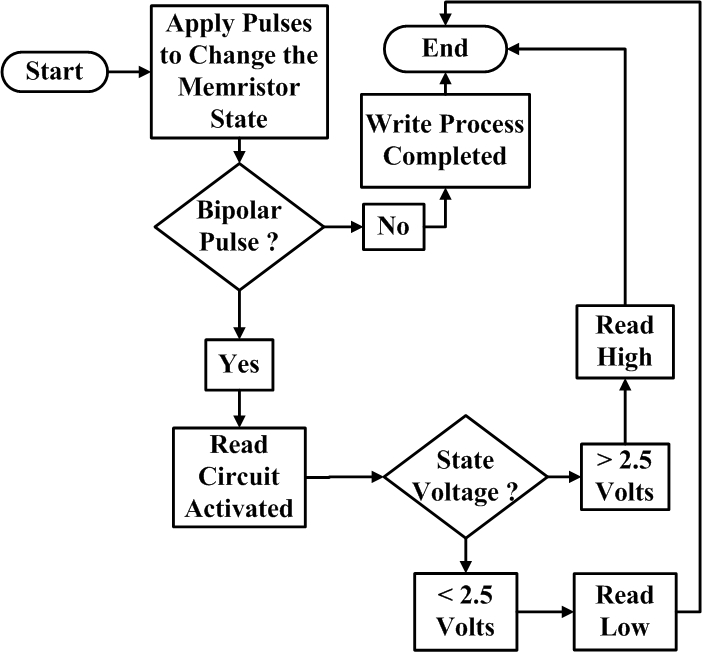}
\caption{\textbf{General Idea Depicting Operation of the Read Write Circuit.}}
\label{Fig: flowchart}
\end{figure}

Flow chart in  Fig. \ref{Fig: flowchart} depicts the general operation of read write circuit. The detailed circuit diagram of proposed memristor emulator based read write circuit is shown in Fig. \ref{Fig: memristorreadwrite2}. The circuit consists of a PMOS transistor, a resistive attenuator, and two high gain operational amplifiers. The circuit is developed based on Fig. \ref{Fig: signalflow} and Fig. \ref{Fig: flowchart}. Q1 and first comparator serve the purpose of detecting bipolar pulse and activating read circuit.
Resistive attenuator is formed by two 1 kilo ohms
resistors. The final comparator compares memristor current state with 2.5 volts and gives single bit output (high or low). 
One of the unique features of this circuit is that it does not need external read write enable switch to toggle between read write operations, instead this is achieved by unipolar write pulse and bipolar read pulse.

\subsection*{Analysis of the Proposed Read Write Circuit}

\begin{figure}[t]
\center
\includegraphics[scale=0.4]{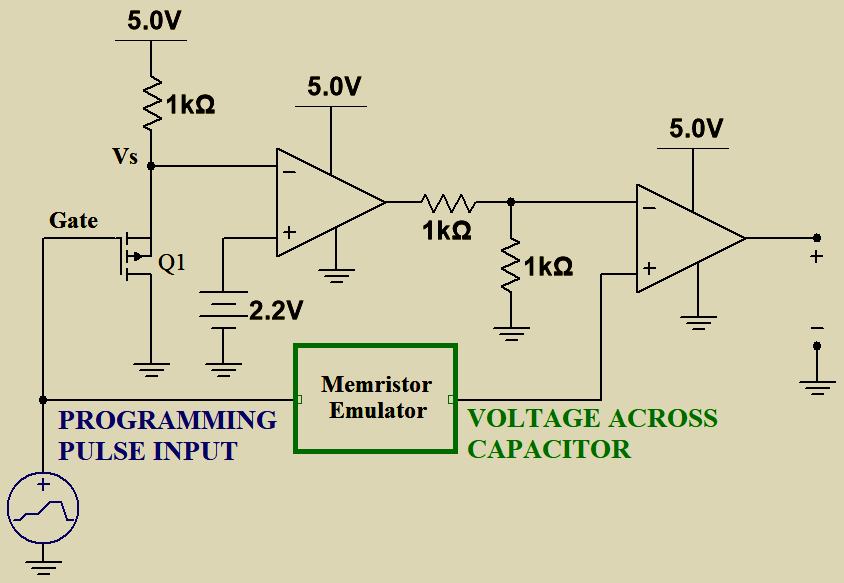}
\caption{\textbf{Proposed Read Write Circuit.}}
\label{Fig: memristorreadwrite2}
\end{figure}

The PMOS transistor in the circuit of Fig. \ref{Fig: memristorreadwrite2}, Q1, has same specifications as the specifications of the transistors used for developing memristor emulator. The threshold voltage ($V_t$) is -0.5 volt and device constant (K) is 1 mA/V$^2$. The gate voltage ($V_G$) becomes positive with respect to ground during entire write cycle due to positive unipolar nature of write pulses. Since the read pulse is bipolar, the read circuit activates only during negative interval of the read cycle.     

When the input pulse is positive, for example +1 V, the transistor remains in  saturation region. The source potential ($V_S$) is computed as \cite{razavi},     
$$
(5-V_S) = \frac{1}{2}K[V_{SG}-|v_t|]^2
$$
$$
(10-2\times V_S) = [V_S-1-0.5]^2
$$ 
$$
V_S = 3.32 V
$$ 
When the input pulse is negative, for example, -1 V volt, the transistor enters into the triode region. 
Neglecting channel length modulation effect, the source voltage is computed as, 
$$
(5-V_S) = K\left[(V_{SG}-|v_t|)\times V_{SD}-\frac{V_{SD}^2}{2}\right]
$$
$$
(5-V_S) = [(V_S+1-0.5)\times V_S-0.5\times V_S^2]
$$ 
$$
V_S = 2 V
$$ 
If $V_S$ is above 2 volts, memristor is in write mode and vice versa. Leaving 0.2 volts as safety margin, the source voltage is compared with 2.2 volts to detect read pulse and read duration. 

\begin{equation}
  {\text{Memristor Mode }}=\left\{
  \begin{array}{@{}ll@{}}
    \text{Write}, & \text{if}\ V_S > 2.2 \text{ V}\\
    \text{Read}, & \text{if}\ V_S < 2.2 \text{ V}
  \end{array}\right.
\end{equation} 
The output of first comparator goes high only when the read pulse is detected; this leaves negative terminal of second comparator at 2.5 volts. By comparing memristor state with this 2.5 volts, final output is taken as the state of memristor from the second comparator.

\section{Experiments and Simulations}

\begin{figure}[t]
\begin{subfigure}{.5\textwidth}
  \centering
  \includegraphics[width=\linewidth]{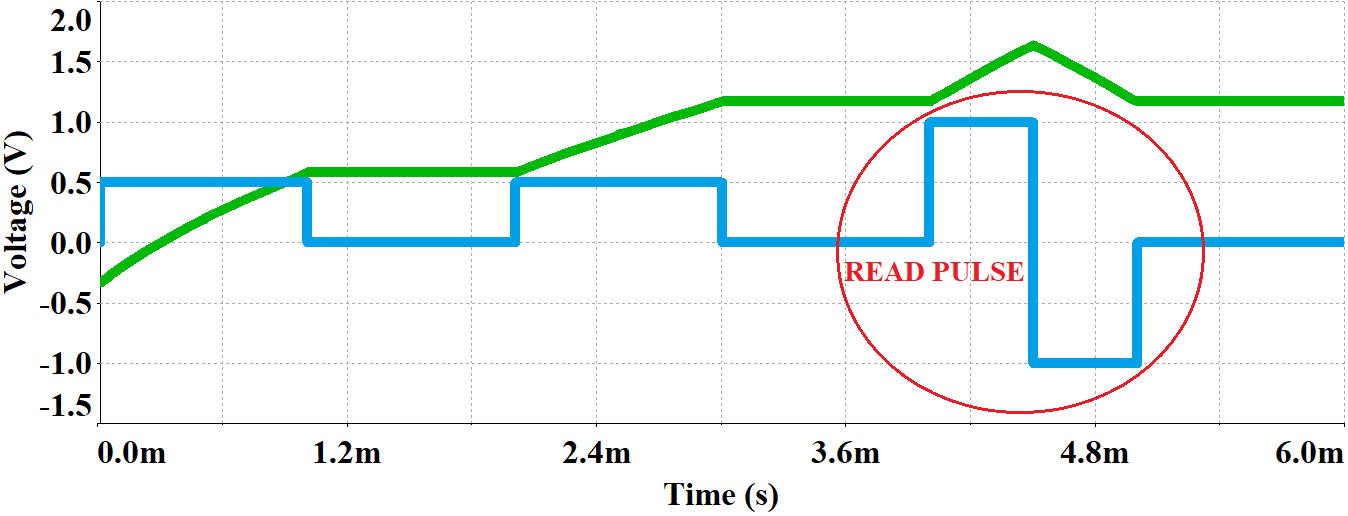}  
  \caption{Returning of memristor to its original state with zero average bipolar read pulse.}
  \label{fig:7a}
\end{subfigure}
\hfill
\begin{subfigure}{.5\textwidth}
  \centering
  \includegraphics[width=\linewidth]{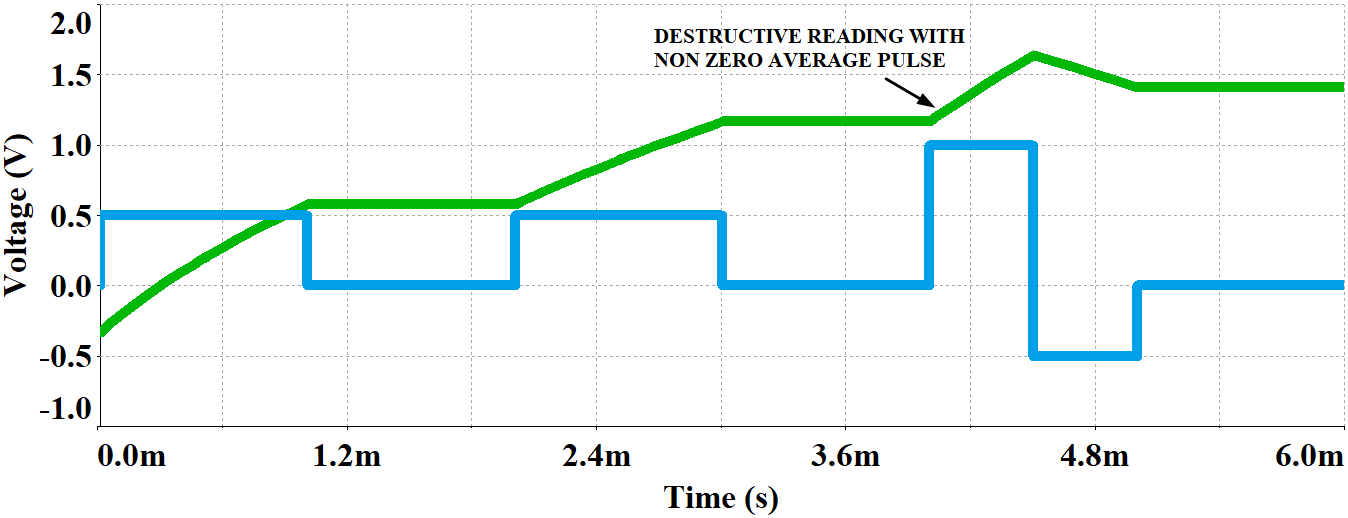}  
  \caption{Memristor fails to return to its original state with non zero average bipolar read pulse.}
  \label{fig:7b}
\end{subfigure}

\caption{\textbf{Effect of Zero Average and Non-zero Average Pulse on the State of the Memristor.}}
\label{Fig: suitablereadpulse}
\end{figure}

A read write circuit is designed using memristor emulator based on the linear memristor model. 
Internal state of the memristor should not be disturbed by the read cycle; it shall return to its original state after the completion of read cycle. 
Simulation has been performed on the read write circuit by applying zero average and non-zero average read pulse. From the simulation results, it is seen that in zero average pulse the memristor state returns to its original state, which does not happen with non-zero average pulse. In Fig. \ref{Fig: suitablereadpulse}(a), memristor returns to it's original state of 1.2 volts after the application of zero average pulse; whereas, memristor fails to return to its original state with non-zero average pulse as shown in Fig. \ref{Fig: suitablereadpulse}(b). Therefore, zero average bipolar pulse is chosen as a read pulse.       


\begin{figure}[t]
\begin{subfigure}{.5\textwidth}
  \centering
  \includegraphics[width=\linewidth]{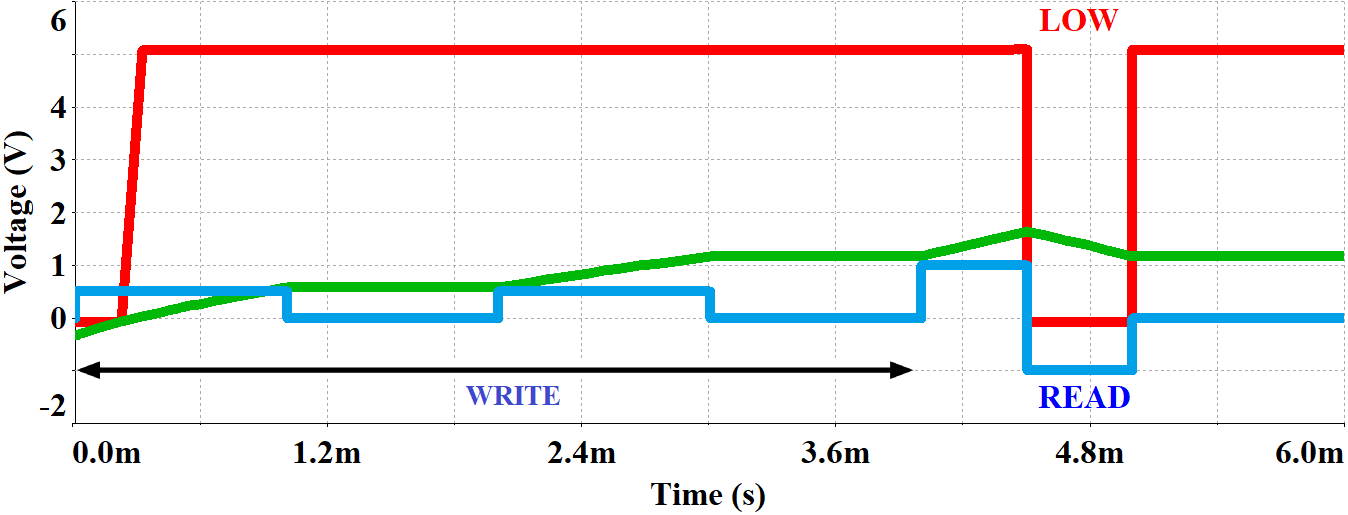}  
  \caption{Reading low by the read circuit. The state of the memristor is below 2.5 volts.}
  \label{fig:8a}
\end{subfigure}
\hfill
\begin{subfigure}{.5\textwidth}
  \centering
  \includegraphics[width=\linewidth]{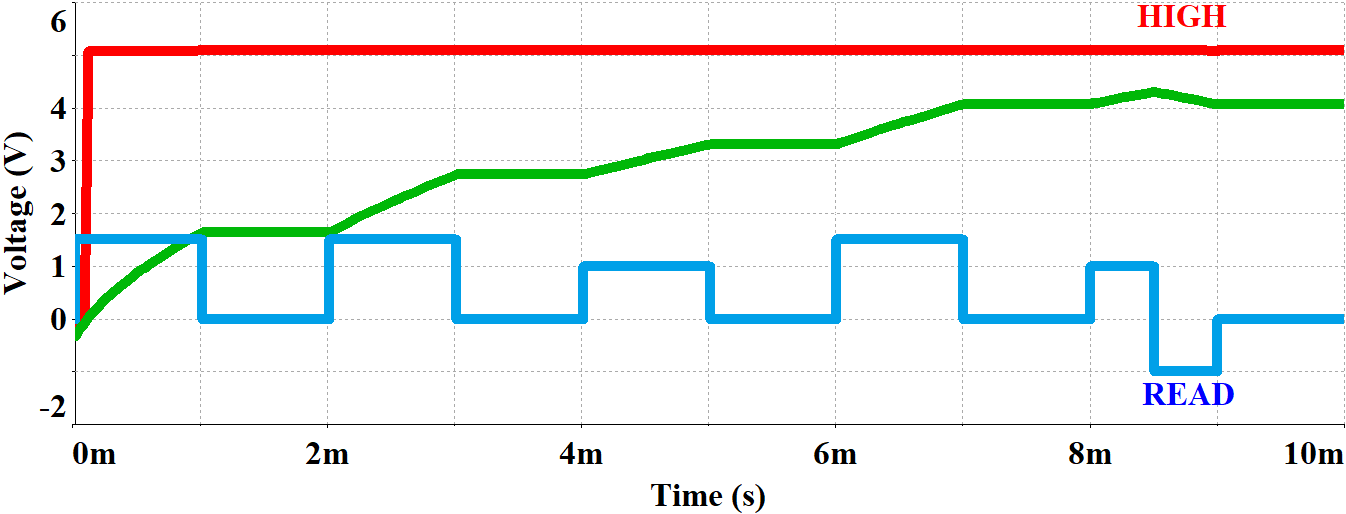}  
  \caption{Reading high by the read circuit.The state of the memristor is above 2.5 volts.}
  \label{fig:8b}
\end{subfigure}

\caption{\textbf{Read Circuit Reading Low and High Memristor State During Negative Cycle of the Bipolar Read Pulse.}}
\label{Fig: readduringreadcycle}
\end{figure}

If the state of the memristor is below 2.5 volts, the output of the read circuit is zero for the read duration. Similarly, if it is above 2.5 volts, the output of the read circuit is 5 volts. 
The complete response of the memristor emulator based read write circuit is shown in Fig. \ref{Fig: readduringreadcycle}. Originally, the output of the read circuit is 5 volts. When the low state is detected, it jumps to the low state (0 volt) as indicated in the Fig. \ref{Fig: readduringreadcycle}(a). If high state is detected during the read cycle,  output of the read circuit does not change its original state as indicated in the Fig. \ref{Fig: readduringreadcycle}(b). Green trace represents memristor state voltage. Blue trace represents programming and read pulse. Red trace represents logic generated by the read write circuit.


\begin{figure}[t]
\center
\includegraphics[width=\linewidth]{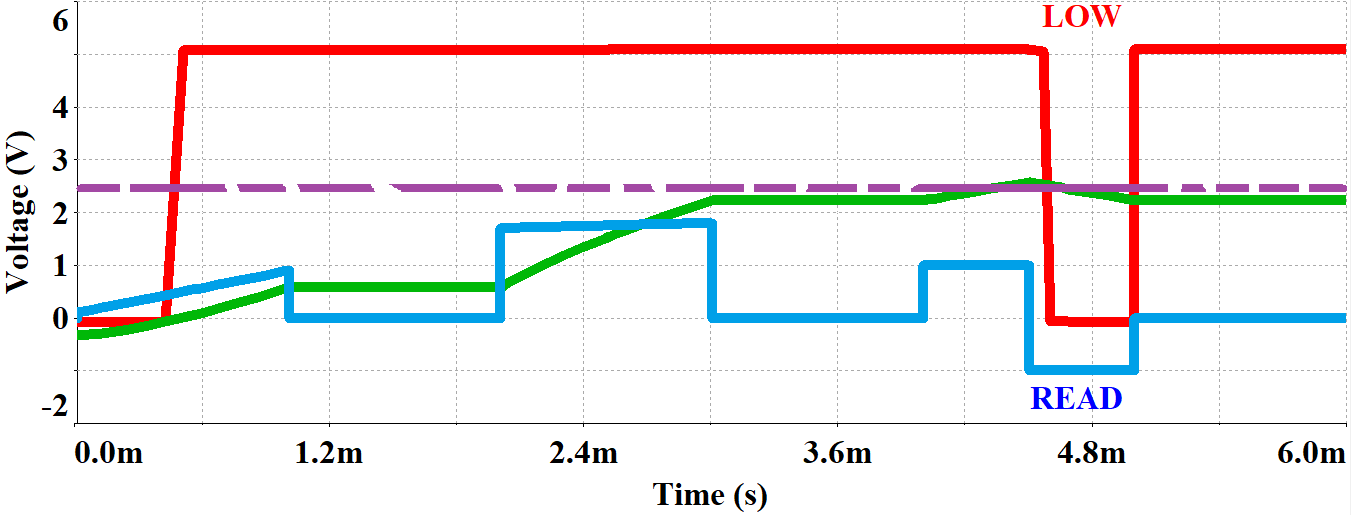}
\caption{\textbf{Reading at the Negative State of Read Pulse to Prevent False Reading Due to Read Cycle Distortion.}}
\label{Fig: dataintegrity}
\end{figure}

During read cycle, the state of memristor changes; however, at the end of read cycle the memristor should return to its original state. The read duration is deliberately chosen as negative state of the read pulse. Due to read cycle distortion, the memristor may enter into incorrect state. Therefore, in order to preserve the data integrity, state of the memristor is 
read at the negative half of the read pulse. As shown in Fig. \ref{Fig: dataintegrity}, due to the read cycle distortion, the state jumps above 2.5 volts (falls on the false region), but the state is correctly read as low during negative state of bipolar read pulse presserving data integrity. Therefore, the designed read circuit is not only simple and non volatile but also resilient to read cycle distortion.  

\section{Conclusion}
The application of memristor as a memory element has been explored through a memristor emulator based memory cell. In this paper, we have designed and developed a read write circuit based on memristor emulator, which can read a single bit. Similarly, the circuit can be taken to any states between 0 to 5 volts representing different memory states. With unipolar write pulse and bipolar read pulse, the circuit can be switched between read and write operations without using an external read/write enable switch.

The read write operations of the proposed circuit are analyzed to ensure the non volatility of the memristor states. Furthermore, the possible read distortion is analyzed and the read access time duration in read pattern is proposed to preserve data integrity during the read cycle.

\ifCLASSOPTIONcaptionsoff
  \newpage
\fi

\end{document}